\documentstyle[11pt,a4]{article}
\input tcilatex
\newtheorem{theorem}{Theorem}[section]
\newtheorem{lemma}[theorem]{Lemma}
\newtheorem{proposition}[theorem]{Proposition}

\QQQ{Language}{
American English
}

\begin{document}

\title{The Feynman Integrand for the Perturbed\\Harmonic Oscillator as a Hida
Distribution}
\author{M\'ario Cunha$^2$ \\Cust\'odia Drumond$^2$\\Peter Leukert$^1$\\Jos\'e Lu\'\i
s Silva$^2$\\Werner Westerkamp$^1$ \medskip\  \\ 
$^1$BiBoS\thanks{%
Research Center Bielefeld-Bochum-Stochastik} -Universit\"at Bielefeld, D
33615 Bielefeld, Germany\\ $^2$Universidade da Madeira, P-9000 Funchal,
Portugal}
\date{April 21, 1994
\thanks{published in Ann. Physik 4 (1995) 53 - 67}}
\maketitle

\begin{abstract}
We review some basic notions and results of White Noise Analysis that are
used in the construction of the Feynman integrand as a generalized White
Noise functional. We show that the Feynman integrand for the harmonic
oscillator in an external potential is a Hida distribution.
\end{abstract}


\section{Introduction}

Path integrals are a useful tool in many branches of theoretical physics
including quantum mechanics, quantum field theory and polymer physics. We
are interested in a rigorous treatment of such path integrals. As our basic
example we think of a quantum mechanical particle.

On one hand it is possible to represent solutions of the heat equation by a
path integral representation, based on the Wiener measure in a
mathematically rigorous way. This is stated by the famous Feynman Kac
formula. On the other hand there have been a lot of attempts to write
solutions of the Schr\"odinger equation as a Feynman (path) integral in a
useful mathematical sense. The methods are always more involved and less
direct than in the euclidean (i.e.\ Feynman Kac) case. Among them are
analytic continuation, limits of finite dimensional approximations and
Fourier transform. We are not interested in giving full reference on various
theories of Feynman integrals (a brief survey can be found in \cite{Ex}) but
we like to mention the method in \cite{AHK} using Fresnel integrals. Here we
have chosen a white noise approach.

White noise analysis is a mathematical framework which offers various
generalizations of concepts known from finite dimensional analysis, among
them are differential operators and Fourier transform. Although we will give
a brief introduction to white noise calculus in section 2 the reader
unfamiliar with this topic is recommended to the monographs \cite{HKPS}, 
\cite{Ob}, \cite{H} and the introductory articles \cite{Kuo}, \cite{P}, \cite
{S2}, \cite{W}.

The idea of realizing Feynman integrals within the White Noise framework
goes back to \cite{HS}. The ''average over all paths'' is performed with a
Hida distribution as the weight (instead of a measure). The existence of
such Hida distributions corresponding to Feynman integrands has been
established in \cite{FPS}. In \cite{KS} Khandekar and Streit moved beyond
the existence theorem by giving an explicit construction for a large class
of potentials including singular ones. Basically they constructed a strong
Dyson series for the Feynman integrand in the space of Hida Distributions.
This approach only works for one space dimension. Their construction was
generalized to (one dimensional) time-dependent potentials of noncompact
support in \cite{LLSW}.

In this work, which originated in the White Noise workshop on Madeira in
August 1993, we carry those ideas over to perturbations of the harmonic
oscillator. Hence instead of constructing a Dyson series around the free
particle Feynman integrand we expand around the Feynman integrand of the
harmonic oscillator as obtained in \cite{FPS}. The external potentials to
which the oscillator is submitted correspond to the wide class of
time-dependent singular potentials treated in \cite{LLSW}.

In \cite[chap 5]{AHK} the path integral of the anharmonic oscillator is
defined within the theory of Fresnel integrals. Compared to our ansatz this
procedure has the advantage of being manifestly independent of the space
dimension. Despite the lack of a generalization to higher dimensional
quantum systems our construction has some interesting features:

\begin{itemize}
\item  The admissible potentials may be very singular.

\item  We are not restricted to smooth initial wave functions and may thus
study the propagator directly.

\item  Instead of giving a meaning to the Feynman {\em integral} we define
the Feynman {\em integrand }as a Hida distribution. By taking expectation we
get the propagator. On the other hand one may now use the toolbox of white
noise analysis and apply differential operators to derive variational
relations or Ehrenfest's theorem, see \cite[chap 12]{HKPS}, \cite{S1}.
\end{itemize}

\section{White Noise Analysis}

\noindent The starting-point of White Noise Analysis is the real Gel`fand
triple\ 

$$
{\cal S}\left( {\bf R}\right) \subset L^2\left( {\bf R}\right) \subset {\cal %
S}^{\prime }\left( {\bf R}\right) , 
$$
where ${\cal S}^{\prime }\left( {\bf R}\right) $ denotes the real Schwartz
space. Using Minlos' theorem we construct the White Noise measure space $%
\left( {\cal S}^{\prime }\left( {\bf R}\right) ,{\cal B},\mu \right) $ by
fixing the characteristic functional in the following way:\ 
$$
C\left( f\right) =\int_{{\cal S}^{\prime }}\exp i\left\langle \omega
,f\right\rangle \text{ }d\mu \left( \omega \right) =\exp \left( -\frac
12\int_{{\bf R}}f^2\left( \tau \right) \;d\tau \right) \text{ },\text{ }f\in 
{\cal S}\left( {\bf R}\right) \ . 
$$
We denote by $\left\langle \cdot ,\cdot \right\rangle $ the bilinear pairing
between ${\cal S}^{\prime }\left( {\bf R}\right) $ and ${\cal S}\left( {\bf R%
}\right) $ and by $\left| \cdot \right| _0$ the norm on $L^2\left( {\bf R}%
\right) $.

\noindent \TeXButton{sloppy}{\sloppy}Within this formalism a version of
Wiener's Brownian motion is given by\ 
$$
B\left( t\right) :=\left\langle \omega ,{\bf 1}_{\left[ 0,t\right)
}\right\rangle =\int_0^t\omega \left( s\right) \text{\thinspace }ds\text{ .} 
$$
We now consider the space $\left( L^2\right) $, which is defined to be the
complex Hilbert space $L^2\left( {\cal S}^{\prime }\left( {\bf R}\right) ,%
{\cal B},\mu \right) $. For applications the space $\left( L^2\right) $ is
often too small. A convenient way to solve this problem is to introduce a
space of test functionals in $\left( L^2\right) $ and to use its larger dual
space.

\TeXButton{fussy}{\fussy}We like to work with the space of test functions $%
\left( {\cal S}\right) .$ So we review the standard construction of $\left( 
{\cal S}\right) $ due to \cite{KT}. For a more detailed discussion see \cite
{HKPS}, \cite{KLPSW}. Take one system of Hilbertian norms $\left\{ \left|
\cdot \right| _p\right\} $ topologizing ${\cal S}\left( {\bf R}\right) $
which grows sufficiently fast. Then ${\cal S}\left( {\bf R}\right) \,$ is
realized as a projective limit of Hilbert spaces ${\cal S}_p\left( {\bf R}%
\right) :$%
$$
{\cal S}\left( {\bf R}\right) =\bigcap\limits_{p\geq 0}{\cal S}_p\left( {\bf %
R}\right) \ , 
$$
where ${\cal S}_p\left( {\bf R}\right) $ denotes the completion of ${\cal S}%
\left( {\bf R}\right) \,$ with respect to $\left| \cdot \right| _p$. Then
the space of tempered distributions is%
$$
{\cal S}^{\prime }\left( {\bf R}\right) =\bigcup\limits_{p\geq 0}{\cal S}%
_{-p}\left( {\bf R}\right) \ , 
$$
where the dual norm $\left| \cdot \right| _{-p}$ topologizes the Hilbert
space ${\cal S}_{-p}\left( {\bf R}\right) $.

One convenient choice is 
\begin{equation}
\label{A}\left| f\right| _p:=\left| A^pf\right| _0,\quad f\in {\cal S}( {\bf %
R}) , 
\end{equation}
where%
$$
Af(t)=-f^{\prime \prime }(t)+\left( t^2+1\right) f(t) 
$$
is the Hamiltonian of the harmonic oscillator. Since $\left( L^2\right) $ is
Segal isomorphic to the complex symmetric Fock space $\Gamma (L^2)$ of $%
L^2\left( {\bf R}\right) $, we can identify the Fock space $\Gamma ({\cal S}%
_p)$ with a subspace $({\cal S})_p$ of $\left( L^2\right) $ and define the
nuclear space%
$$
\left( {\cal S}\right) =\bigcap_{p\geq 0}\left( {\cal S}\right) _p\ . 
$$
Thus we arrive at the Gel'fand triple:\ 
$$
\left( {\cal S}\right) \subset \left( L^2\right) \subset \left( {\cal S}%
\right) ^{*}. 
$$
Elements of the space $\left( {\cal S}\right) ^{*}$ are called Hida
distributions (or generalized Brownian functionals). It is possible to
characterize the spaces $\left( {\cal S}\right) $ and $\left( {\cal S}%
\right) ^{*}$ by their $S$- or $T$-transforms $\left( \Phi \in \left( {\cal S%
}\right) ^{*},\text{ }f\in {\cal S}\left( {\bf R}\right) \right) :$%
\begin{equation}
\label{Tee}T\Phi \left( f\right) \equiv \left\langle \!\left\langle \Phi
,\exp \left( i\left\langle \cdot ,f\right\rangle \right) \right\rangle
\!\right\rangle =\dint_{{\cal S}^{\prime }\left( {\bf R}\right) }\exp \left(
i\left\langle \omega ,f\right\rangle \right) \Phi \left( \omega \right) d\mu
\left( \omega \right) , 
\end{equation}
$$
S\Phi \left( f\right) \equiv \left\langle \!\left\langle \Phi ,:\exp
\left\langle \cdot ,f\right\rangle :\right\rangle \!\right\rangle , 
$$
here $\left\langle \!\left\langle \cdot ,\cdot \right\rangle \!\right\rangle 
$ denotes the bilinear pairing between $\left( {\cal S}\right) $ and $\left( 
{\cal S}\right) ^{*}$ and we have used the traditional notation:\ 

\begin{equation}
\label{Wick}:\exp \left\langle \cdot ,f\right\rangle :\text{ }\equiv C\left(
f\right) \exp \left( \left\langle \cdot ,f\right\rangle \right) ,\text{ }%
f\in {\cal S}({\bf R})\text{ .} 
\end{equation}
We denote by ${\bf E}\left( \Phi \right) \equiv \left\langle \!\left\langle
\Phi ,1\right\rangle \!\right\rangle $ the expectation of a Hida
distribution $\Phi .$ $S$- and $T$-transform have extensions to the complex
Schwartz space ${\cal S}_{{\bf C}}\left( {\bf R}\right) $ and are related by
the following formula: 
\begin{equation}
\label{strafo}S\Phi \left( f\right) =C\left( f\right) \text{ }T\Phi \left(
-if\right) ,\text{ }f\in {\cal S}_{{\bf C}}\left( {\bf R}\right) \text{ } 
\end{equation}
Let us now quote the above mentioned characterization theorem, which is due
to Potthoff and Streit \cite{PS} and has been generalized in various ways
(see e.g. \cite{KoS}, \cite{MY}, \cite{SW}). For a full proof of a
generalized version see \cite{KLPSW}.\ \medskip\ 

\noindent {\bf Theorem 2.1}\smallskip 

\noindent {\it The following statements are equivalent}:\ 

\begin{enumerate}
\item  $F:$ ${\cal S}({\bf R)}\rightarrow {\bf C}$ {\it is }\ 

{\it (A)\quad {\bf Ray-entire}, i.e. for all} $g,f\in {\cal S}\left( {\bf R}%
\right) $ {\it the mapping }${\bf C\ni }$ $z\mapsto F(zf+g)$ {\it is entire}$%
.$\ 

{\it (B)\quad and uniformly of order two, i.e. there exist constants} $%
K_1,K_2>0$ {\it such that }%
$$
\left| F\left( zf\right) \right| \leq K_1\exp \left( K_2\left| z\right|
^2\left| f\right| ^2\right) ,\qquad f\in {\cal S}\left( {\bf R}\right) . 
$$
\TeXButton{10mm}{\hspace*{10mm}}{\it for some continuous norm }$\left| \cdot
\right| ${\it \ on }${\cal S}\left( {\bf R}\right) .$

\item  $F$ {\it is the} $S${\it - transform of a unique Hida distribution} $%
\Phi \in \left( {\cal S}\right) ^{*}.$\ 

\item  $F$ {\it is the }$T${\it - transform of a unique Hida distribution} $%
\stackrel{\wedge }{\Phi }$ $\in \left( {\cal S}\right) ^{*}.$\ 
\end{enumerate}

\noindent A functional satisfying 1. is usually called a $U$-functional.\ 

As an example of an application of this theorem we consider Donsker`s delta
function.

\noindent Consider the composition $\delta _a\circ B\left( t\right) $ of the
Dirac distribution $\delta _a$ at $a\in {\bf R}$ with Brownian motion $B(t)$%
, $t>0$:\ 
$$
\Phi =\delta \left( B\left( t\right) -a\right) 
$$
\begin{equation}
\label{Donsker}\Phi =\delta \left( \left\langle \cdot ,{\bf 1}_{\left[
0,t\right) }\right\rangle -a\right) ,\text{ }a\in {\bf R}.\text{ } 
\end{equation}
\ The $S$-transform of $\Phi $ is calculated to be \cite{HKPS}:%
$$
S\Phi \left( f\right) =\frac 1{\sqrt{2\pi t}}\exp \left( -\frac 1{2t}\left(
\int_0^tf\left( s\right) \text{\thinspace }ds-a\right) ^2\right) 
$$
and theorem 2.1 gives immediately that $\Phi $ is a well defined element in $%
\left( {\cal S}\right) ^{*}$.\medskip\ \ 

Now we want to mention some important consequences of theorem 2.1. The first
one concerns the convergence of sequences of Hida distributions and can be
found in \cite{HKPS}, \cite{PS}, \cite{KLPSW}.\medskip\ 

\noindent {\bf Theorem 2.2}\smallskip\ 

\noindent {\it Let} $\left\{ F_n\right\} _{n\in {\bf N}}$ {\it denote a
sequence of }$U${\it -functionals with the following properties:}

\begin{enumerate}
\item  {\it For all} $f\in {\cal S}\left( {\bf R}\right) $ , $\left\{
F_n\left( f\right) \right\} _{n\in {\bf N}}$ {\it is a Cauchy sequence,}\ 

\item  {\it There exist} $K_{1,}$ $K_2$ $>0$ {\it such that the bound}\ 
$$
\left| F_n\left( zf\right) \right| \leq K_1\exp \left( K_2\left| z\right|
^2\left| f\right| ^2\right) \text{ },\text{ }f\in {\cal S}\left( {\bf R}%
\right) 
$$
{\it holds for almost all }$n\in {\bf N}$ {\it in a continuous norm }$\left|
\cdot \right| ${\it \ on }${\cal S}\left( {\bf R}\right) ${\it . }
\end{enumerate}

\noindent {\it Then there is a unique} $\Phi \in \left( {\cal S}\right) ^{*}$
{\it such that }$T^{-1}F_n$ {\it converges strongly to }$\Phi .$\ \medskip\
\ \ 

\noindent This theorem is also valid for $S$-transforms.\smallskip\ 

Another corollary of theorem 2.1 deals with the integration of Hida
distributions which depend on an additional parameter (see \cite{HKPS}, \cite
{KS}, \cite{KLPSW}).\bigskip
\ \ \ 

\noindent {\bf Theorem 2.3}\smallskip\ 

\noindent {\it Let} $\left( \Omega ,B,m\right) $ {\it denote a measure space
and} $\lambda \mapsto \Phi \left( \lambda \right) $ {\it a mapping from} $%
\Omega ${\it \ to} $\left( {\cal S}\right) ^{*}$. {\it Let }$F\left( \lambda
\right) $ {\it denote the} $T$-{\it transform of }$\Phi \left( \lambda
\right) $ {\it which satisfies the following conditions:}\ 

\begin{enumerate}
\item  $\lambda \mapsto F\left( \lambda ,f\right) $ {\it is a measurable
function for all} $f\in {\cal S}\left( {\bf R}\right) ,$\ 

\item  {\it There exists} {\it a continuous norm }$\left| \cdot \right| $%
{\it \ on }${\cal S}\left( {\bf R}\right) $ {\it such that}%
$$
\ \left| F\text{ }(\lambda ,zf)\right| \leq K_1\left( \lambda \right) \exp
\left( K_2\left( \lambda \right) \left| z\right| ^2\left| f\right| ^2\right)
,\text{ \quad }f\in {\cal S}\left( {\bf R}\right) 
$$
{\it with} $K_1$ $\in $ $L^1\left( \Omega ,m\right) $ {\it and} $K_2\in
L^\infty \left( \Omega ,m\right) .$
\end{enumerate}

\noindent {\it Then} $\Phi $ {\it is Bochner integrable in some }$\left( 
{\cal S}\right) _{-q}$ {\it and thus}\ 
$$
\int\limits_\Omega \Phi \left( \lambda \right) dm\left( \lambda \right) \in
\left( {\cal S}\right) ^{*}. 
$$
$T$-{\it transform and integration commute}\ 
$$
T\left( \int\limits_\Omega \Phi \left( \lambda \right) \text{ }dm\left(
\lambda \right) \right) \left( f\right) =\int\limits_\Omega T\left( \Phi
\left( \lambda \right) \right) \left( f\right) \,dm\left( \lambda \right) . 
$$
\noindent Again the same theorem holds for the $S$-transform.\bigskip 

\noindent {\bf Example }%
$$
\delta (B(t)-a)=\frac 1{2\pi }\int_{{\bf R}}e^{i\lambda (B(t)-a)}d\lambda 
$$
in the sense of Bochner integration (see, e.g., \cite{HKPS}).\bigskip\ \ 

\noindent {\bf Remark}

For later use we have to define pointwise products of a Hida distribution $%
\Phi $ with a Donsker-Delta function%
$$
\delta \left( \langle \omega ,g\rangle -a\right) 
$$
If the mapping $\lambda \longmapsto T\Phi (f+\lambda g)$ is integrable on%
{\bf \ R} the following formula may be used to define the product $\Phi
\cdot \delta $%
\begin{equation}
\label{phidelta}T\left( \Phi \cdot \delta (\langle \omega ,g\rangle
-a)\right) (f)=\frac 1{2\pi }\int_{{\bf R}}e^{-i\lambda a}\ T\Phi \left(
f+\lambda g\right) \ d\lambda \text{,} 
\end{equation}
in case the right hand integral is indeed a U-functional.

\section{The Feynman Integrand as a Hida Distribution}

We follow \cite{FPS} and \cite{HS} in viewing the Feynman integral as a
weighted average over Brownian paths. These paths are modeled within the
White Noise framework according to\smallskip\ 
$$
x\left( t\right) \equiv x_0+\sqrt{\frac \hbar m}\int_{t_0}^t\omega \left(
\tau \right) \text{ }d\tau , 
$$
in the sequel we set $\hbar =m=1.$\smallskip\ 

\noindent In \cite{FPS} the (distribution-valued) weight for the free
quantum mechanical propagation from $x\left( t_0\right) =x_0$ to $x\left(
t\right) =x$ is constructed from a kinetic energy factor $\exp \left( \frac
i2\int_{t_0}^t\omega ^2\left( \tau \right) \text{ }d\tau \right) $ and a
Donsker delta function $\delta (x\left( t\right) -x)$. Furthermore a factor $%
\exp \left( \frac 12\text{ }\int_{t_0}^t\omega ^2\left( \tau \right) \text{ }%
d\tau \right) $ is introduced to compensate the Gaussian fall-off of the
White Noise measure in order to mimic Feynman's non-existing ''flat''
measure $D^\infty x.$ Thus in \cite{FPS} the Feynman integrand for the free
motion reads (the N indicates appropriate normalization):\smallskip\ 
$$
I_0={\rm Nexp}\left( \frac{i+1}2\int_{t_0}^t\omega ^2\left( \tau \right) 
\text{ }d\tau \right) \delta \left( x\left( t\right) -x\right) \text{.} 
$$
As in \cite{FPS} $I_0$ is a Hida distribution, with $T$- transform given by\ 
$$
TI_0\left( f\right) =\frac 1{\sqrt{2\pi i\left| t-t_0\right| }}\exp \left(
-\frac i2\left| f_\Delta \right| ^2-\frac 12\left| f_{\Delta ^c}\right|
^2+\frac i{2\left| t-t_0\right| }\left( \int_{t_0}^tf\left( \tau \right)
d\tau +x-x_0\right) ^2\right) \text{,} 
$$
where $\Delta =\left[ t_0,t\right] $ and $f_\Delta $, $f_{\Delta ^c}$ denote
the restrictions of $f$ to $\Delta $ and its complement $\Delta ^c$
respectively. Furthermore the Feynman integral ${\bf E}\left( I_0\right)
=TI_0\left( 0\right) $ is indeed the free particle propagator $\frac 1{\sqrt{%
2\pi i\left| t-t_0\right| }}$ $\exp \left[ \frac i{2\left| t-t_0\right| }%
\text{ }\left( x-x_0\right) ^2\right] $. Not only the expectation but also
the $T$- transform has a physical meaning. By a formal integration by parts\ 
$$
TI_0\left( f\right) ={\bf E}\Big( I_0\text{ }e\text{ }^{-i\int_{t_0}^tx%
\left( \tau \right) \stackrel{\cdot }{f{}}\left( \tau \right) \,d\tau }%
\Big)
\text{ }e\text{ }^{ixf\left( t\right) -ix_0f\left( t_0\right) }\text{ }e%
\text{ }^{-\frac 12\left| f_\Delta c\right| ^2}. 
$$
The term $e$ $^{-i\int_{t_0}^tx\left( \tau \right) \stackrel{\cdot }{f{}}%
\left( \tau \right) \,d\tau }$ would thus arise from a time-dependent
potential $W\left( x,t\right) =$ $\dot f{}{}(\tau )x$. And indeed it is
straightforward to verify that\ 
\begin{equation}
\label{KoDef}\theta \left( t-t_0\right) TI_0\left( f\right) =K_0^{\left(
\dot f\right) {}}\left( x,t|x_0,t_0\right) \text{ }e\text{ }^{ixf\left(
t\right) -ix_0f\left( t_0\right) }\text{ }e\text{ }^{-\frac 12\left|
f_\Delta c\right| ^2},
\end{equation}
where%
$$
K_0^{\left( \dot f\right) }\left( x,t|x_0,t_0\right) =\frac{\theta \left(
t-t_0\right) }{\sqrt{2\pi i\left| t-t_0\right| }}\;\times  
$$
$$
\exp \left( ix_0f\left( t_0\right) -ixf\left( t\right) -\frac i2\left|
f_\Delta \right| ^2+\frac i{2\left| t-t_0\right| }\left( \int_{t_0}^tf\left(
\tau \right) d\tau +x-x_0\right) ^2\right)  
$$
is the Green's function corresponding to the potential $W,$ i.e. $%
K_0^{\left( \dot f\right) }$ obeys the Schr\"odinger equation\ 
$$
\left( i\partial _t+\frac 12\partial _x^2-\dot f{}\left( t\right) x\right)
K_0^{\left( \dot f\right) }\left( x,t|x_0,t_0\right) =i\,\delta \left(
t-t_0\right) \,\delta \left( x-x_0\right) . 
$$
More generally one calculates\ 
\begin{equation}
\label{Tnd}\theta (t-t_0)T\left( I_0\stackrel{n+1}{\stackunder{i=1}{\tprod }}%
\delta \left( x\left( t_i\right) -x_i\right) \right) \left( f\right)
=e^{-\frac 12\left| f_\Delta c\right| ^2}{}e^{\text{ }^{ixf\left( t\right)
-ix_0f\left( t_0\right) }}\stackrel{n+1}{\stackunder{i=1}{\tprod }}%
K_0^{\left( \dot f\right) }\left( x_i,t_i|x_{i-1},t_{i-1}\right) .
\end{equation}
Here $t_0<t_1<...<t_n<t_{n+1}\equiv t$ and $x_{n+1}\equiv x$ .\medskip\ 

In order to pass from the free motion to more general situations, one has to
give a rigorous definition of the heuristic expression

$$
I=I_0\exp \left( -i\int_{t_0}^tV\left( x\left( \tau \right) \right) \text{ }%
d\tau \right) . 
$$

In \cite{KS} Khandekar and Streit accomplished this by perturbative methods
in case $V$ is a finite signed Borel measure with compact support. This
construction was generalized in \cite{LLSW} to a wider class of potentials
by allowing time-dependent potentials and a Gaussian fall-off instead of a
bounded support.

\noindent The starting point is a power series expansion of $\exp \left(
-i\int_{t_0}^tV\left( x\left( \tau \right) ,\tau \right) d\tau \right) $
using \\$V\left( x\left( \tau \right) ,\tau \right) =\int dx\,V\left( x,\tau
\right) \,\delta \left( x\left( \tau \right) -x\right) :$%
$$
\exp \left( -i\int_{t_0}^tV\left( x\left( \tau \right) ,\tau \right) d\tau
\right) =\dsum\limits_{n=0}^\infty \left( -i\right) ^n\int_{\Lambda _n}d^nt%
\stackrel{n}{\stackunder{i=1}{\tprod }}\int dx_i\,V\left( x_i,t_i\right)
\delta \left( x\left( t_i\right) -x_i\right)  
$$
where $\Lambda _n=\left\{ \left( t_1,...,t_n\right)
|\,t_0<t_1<...<t_n<t\right\} $.\medskip 

In order to consider singular potentials $V$ is no longer taken to be a
function $V$ but a measure $\nu $. Under suitable conditions on $\nu $ it is
proven in \cite{KS} and \cite{LLSW} that%
$$
I_V=I+\sum_{n=1}^\infty \left( -i\right) ^n\int_{{\bf R^{\mit n}}%
}\int_{\Lambda _n}{\textstyle \bigg(\prod\limits_{j=1}^n\nu (dx_j,dt_j)\bigg)%
}I_0\prod_{j=1}^n\delta \left( x\left( t_j\right) -x_j\right) 
$$
exists as a well-defined element of $({\cal S})^{*}$ using theorems 2.2 and
2.3. \ 

\section{\bf The unperturbed harmonic oscillator}

In this section we first review some results of \cite{FPS} which are
necessary for the formulation and proof of our main result. Then we prepare
a proposition on which we base our perturbative method.

To define the Feynman integrand 
$$
I_h=I_0\exp \left( -i\int_{t_0}^tU\left( x(\tau )\right) \ d\tau \right) 
\text{, }U(x)=\frac 12k^2x^2 
$$
of the harmonic oscillator, at least two things have to be done.

First we have to justify the pointwise multiplication of $I_0$ with the
interaction term and secondly it has to be shown that ${\bf E}(I_h)$ solves
the Schr\"odinger equation for the harmonic oscillator. Both has been done
in \cite{FPS}. There the $T$-transform of $I_h$ has been calculated and
shown to be a $U$-Functional. Thus $I_h\in ({\cal S})^{*}$. Later we will
use the following modified version of their result: 
$$
TI_h\left( f\right) =\sqrt{\frac k{2\pi i\sin k\left| \Delta \right| }}\exp
\left( -\frac i2\left| f_\Delta \right| ^2-\frac 12\left| f_{\Delta
^c}\right| ^2\right) \exp \bigg\{ \frac{ik}{2\sin k\left| \Delta \right| }%
\bigg[ \left( x_0^2+x^2\right) \cdot  
$$
$$
\cdot \cos k\left| \Delta \right| -2x_0x+2x\int_{t_0}^tdt^{\prime }f\left(
t^{\prime }\right) \cos k\left( t^{\prime }-t_0\right)
-2x_0\int_{t_0}^tdt^{\prime }f\left( t^{\prime }\right) \cos k\left(
t-t^{\prime }\right) + 
$$
\begin{equation}
\label{17}+2\int_{t_0}^tds_1\int_{t_0}^{s_1}ds_2f\left( s_1\right) f\left(
s_2\right) \cos k\left( t-s_1\right) \cos k\left( s_2-t_0\right) \bigg] 
\bigg\} ,
\end{equation}
with $0<k\left| \Delta \right| <\dfrac \pi 2$ , which is easily seen to be a 
$U$-functional.

For our purposes it is convenient to introduce 
$$
K_h^{\left( \dot f\right) }(x,t\mid x_0,t_0)=\theta (t-t_0)\ TI_h(f)\cdot
\exp {\textstyle\frac 12}\left| f_\Delta c\right| ^2\cdot \exp \left(
ix_0f(t_0)-ixf(t)\right) \text{ ,} 
$$
which is the propagator of a particle in a time dependent potential $\frac
12k^2x^2+x\dot f(t).$ This allows for an independent check on the
correctness of the above result. In advanced textbooks of quantum mechanics
such as \cite{Ho} the propagator for an harmonic oscillator coupled to a
source $j$ (forced harmonic oscillator) is worked out. Upon setting $j=\dot f
$ their result is easily seen to coincide with the formula given above.

Proceeding exactly as in the free case (see \cite{KS}, \cite{LLSW}) we first
have to define the (pointwise) product 
$$
I_h\tprod\limits_{j=1}^n\delta \left( B(t_j)-x_j\right) 
$$
in $({\cal S})^{*}$. The expectation of this object can be interpreted as
the propagator of a particle in a harmonic potential, where the paths all
are ''pinned'' such that $B(t_j)=x_j$ , $1\leq j\leq n$. Following the ideas
of the remark at the end of the section 2 we will have to apply (\ref
{phidelta}) repeatedly. But due to the form of $TI_h(f),$ which contains $f$
only in the exponent up to second order, all these integrals are expected to
be Gaussian.

Using this we arrive at the following

\begin{proposition}
For $x_0<x_j<x$, $1\leq j\leq n,\quad $ $t_0<t_j<t_{j+1}<t$, $1\leq j\leq n-1
$,\\ $\quad I_h\prod_{j=1}^n\delta \left( B(t_j)-x_j\right) $ is a Hida
distribution and its $T$-transform is given by%
$$
T\Big(I_h\tprod\limits_{j=1}^n\delta \left( B(t_j)-x_j\right) \Big)%
(f)=e^{-\frac 12\left| f\Delta c\right| ^2}e^{i\left( xf(t)-x_of(t_o)\right)
}\tprod\limits_{j=1}^{n+1}K_h^{\left( \dot f\right) }\left(
x_{j-1},t_{j-1}|x_j,t_j\right) \text{ .} 
$$
\end{proposition}

\TeXButton{Proof}{\proof}For $n=1$ we may check the assertion by direct
computation using formula (\ref{phidelta}). To perform induction one needs
the following

\begin{lemma}
Let $[t_0,t]\subset [t_0^{^{\prime }},t^{^{\prime }}]$ then%
$$
K_h^{\left( \big(f+\lambda {\bf 1}_{[t_0^{^{\prime }},t^{^{\prime }}]}\big)^{%
\displaystyle \cdot }\right) }\left( x_0,t_0|x,t\right) =K_h^{\left( \dot
f\right) }\left( x_0,t_0|x,t\right) ,\text{ }\forall \lambda \in {\bf R}%
\text{ .} 
$$
\end{lemma}

The Lemma is also proven by a lengthy but straightforward computation. On a
formal level the assertion of the lemma is obvious as both sides of the
equation are solutions of the same Schr\"odinger equation if $[t_0,t]\subset
[t_0^{^{\prime }},t^{^{\prime }}]$ . \TeXButton{End Proof}{\endproof}\bigskip%
\ 

The proposition states what one intuitively expects, ordinary propagation
from one intermediate position to the next.

\section{The Feynman integrand for the harmonic oscillator\protect\\in an
external potential}

In this section we construct the Feynman integrand for the harmonic
oscillator in an external potential V(x,t). Thus we have to define 
$$
I_V=I_h\cdot \exp \left( -i\int_{t_0}^tV(x(\tau ),\tau )d\tau \right) \text{
.} 
$$
As for the free particle we introduce the perturbation $V$ via the series
expansion of the exponential. Hence we have to find conditions for $V$ such
that the following object exists in $\left( {\cal S}\right) ^{*}$ 
$$
I_V=I_h+\sum_{n=1}^\infty \left( -i\right) ^n\int_{{\bf R^{\mit n}}%
}d^nx\int_{\Lambda _n}d^nt\prod_{j=1}^nV\left( x_j,t_j\right) \delta \left(
x\left( t_j\right) -x_j\right) I_h\text{ .} 
$$
Since we want to study singular time-dependent potentials, we consider $\nu $
a finite signed Borel measure on ${\bf R}\times \Delta $. Let $\nu _x$
denote the marginal measure 
$$
\nu _x\left( A\in {\cal B}\left( {\bf R}\right) \right) \equiv \nu \left(
A\times \Delta \right) 
$$
and similarly%
$$
\nu _t\left( B\in {\cal B}\left( \Delta \right) \right) \equiv \nu \left( 
{\bf R}\times B\right) . 
$$
The following theorem contains conditions under which the Feynman integrand $%
I_V$ exists as a Hida distribution.

\begin{theorem}
Let $\nu \equiv \nu _{+}-\nu _{-}$ be a finite signed Borel measure on ${\bf %
R}\times \Delta $ where the marginal measures $|\nu |_x:=(\nu _{+}+\nu
_{-})_x$ and $|\nu |_t$ satisfy\medskip\ 

i) $\exists R>0,$ $\forall r>R:\qquad $ $\left| \nu \right| _x\left( \left\{
x:\left| x\right| >r\right\} \right) <\exp \left( -\beta r^2\right) \quad $
for some $\beta >0;$

ii) $\left| \nu \right| _t$ has a $L^\infty $ density.\medskip\ 

\noindent Then 
\begin{equation}
\label{result}I_V=I_h+\sum_{n=1}^\infty \left( -i\right) ^n\int_{{\bf R^{%
\mit n}}}\int_{\Lambda _n}\bigg(\tprod\limits_{j=1}^n\nu (dx_j,dt_j)\bigg)%
I_h\prod_{j=1}^n\delta \left( x\left( t_j\right) -x_j\right) 
\end{equation}
is a Hida distribution.
\end{theorem}

\noindent {\bf Remark:} Conditions {\it i}) and {\it ii}) allow for some
rather singular potentials, e.g.\ $\sum e^{-n^2}\delta _{_n}$. For a cut-off
interaction, i.e. compactly supported $\nu _x$, condition {\it i}) is of
course valid. Note also that $\nu $ is not supposed to be a product measure,
hence the time dependence can be more intricate than simple multiplication
by a function of time.\bigskip\ 

\noindent {\bf Proof.} \\{\bf 1.\ part:} In the first part of the proof we
have to perform some technicalities which are necessary to establish the
central estimate (\ref{Main}). We have to use a very careful procedure to
achieve that (\ref{Main}) survives $n$-fold integration and summation in the
second part of the proof.

From proposition 4.1 and the explicit formula \ref{17} we find 
$$
\Big|T\Big(I_h\tprod\limits_{j=1}^n\delta \left( B(t_j)-x_j\right) \Big)%
\left( zf\right) \Big|\leq e^{\frac{\left| z\right| ^2}2\left| f\right|
_0^2}\left( {\textstyle \prod\limits_{j=1}^{n+1}\sqrt{\frac 1{4\left| \Delta
_j\right| }}}\right) \exp \left( (\left| x_{n+1}\right| +\left| x_o\right|
)\frac \pi 2\left| z\right| \sup _\Delta \left| f\right| \right) \cdot  
$$
$$
\cdot \bigg|\exp \bigg( \bigg\{ \sum_{j=1}^nikzx_j\Big[ \frac 1{\sin k\left|
\Delta _j\right| }\int_{\Delta _j}dtf\left( t\right) \cos k\left(
t-t_{j-1}\right) \TeXButton{5cm}{\hspace*{5cm}} 
$$
$$
\TeXButton{5cm}{\hspace*{5cm}}-\frac 1{\sin k\left| \Delta _{j+1}\right|
}\int_{\Delta _j+1}\!dtf\left( t\right) \cos k\left( t-t_{j+1}\right) 
\Big] 
\bigg\} \bigg) \bigg|
\cdot  
$$
$$
\cdot \exp \left\{ \sum_{j=1}^n\frac{\pi \left| z\right| ^2}{2\left| \Delta
_j\right| }\int_{\Delta _j}ds_1\int_{\Delta _j}ds_2\left| f\left( s_1\right)
\right| \left| f\left( s_2\right) \right| \right\}  
$$
We define 
$$
X=\sup _{0\leq j\leq n+1}\left| x_j\right|  
$$
and%
$$
\left\| f\right\| \equiv \sup _\Delta \left| f\right| +\sup _\Delta \left|
f^{\prime }\right| +\left| f\right| _o 
$$
With these 
$$
\Big|T\Big(I_h\tprod\limits_{j=1}^n\delta \left( B(t_j)-x_j\right) \Big)%
\left( zf\right) \Big|\leq e^{\frac{\left| z\right| ^2}2\left\| f\right\|
^2}\left( {\textstyle \prod\limits_{j=1}^{n+1}\sqrt{\frac 1{4\left| \Delta
_j\right| }}}\right) \exp \left( X\pi \left| z\right| \left\| f\right\| +%
\frac{\pi \left| z\right| ^2}2\left| \Delta \right| \left\| f\right\|
^2\right) \cdot  
$$
$$
\cdot \bigg| \exp \bigg( \bigg\{ \sum_{j=1}^nikzx_j\Big[ \frac 1{\sin
k\left| \Delta _j\right| }\int_{\Delta _j}dtf\left( t\right) \cos k\left(
t-t_{j-1}\right) \TeXButton{4cm}{\hspace*{4cm}} 
$$
$$
\TeXButton{4cm}{\hspace*{4cm}}-\frac 1{\sin k\left| \Delta _{j+1}\right|
}\int_{\Delta _j+1}\!dtf\left( t\right) \cos k\left( t-t_{j+1}\right) 
\Big] 
\bigg\} \bigg) \bigg| \text{ .} 
$$
To estimate the last factor we proceed as follows:%
$$
\left| \sum_{j=1}^nikzx_j\left[ \frac 1{\sin k\left| \Delta _{j+1}\right|
}\int_{\Delta _{j+1}}\!dtf\left( t\right) \cos k\left( t-t_j\right) -\frac
1{\sin k\left| \Delta _{j+1}\right| }\int_{\Delta _j+1}\!dtf\left( t\right)
\cos k\left( t-t_{j+1}\right) \right] \right|  
$$
\vspace{-5mm}%
\begin{eqnarray*}
&\leq& \sum_{j=1}^nk\left| z\right| X\frac 1{\sin k\left| \Delta _{j+1}\right|
}\left| \int_{\Delta _{j+1}}dtf\left( t\right) \int_{t_j}^{t_{j+1}}k\sin
k\left( t-\tau \right) d\tau \right|\\
&\leq& \sum_{j=1}^nk\left| z\right| X\sup _\Delta \left| f\right| \frac \pi
2\left| \Delta _{j+1}\right|\\
&\leq& \frac \pi 2\left| z\right| Xk\left\| f\right\| \left| \Delta \right|
\end{eqnarray*}
To obtain a bound for the remaining term%
$$
\left| \sum_{j=1}^nikzx_j\left[ \frac 1{\sin k\left| \Delta _j\right|
}\int_{\Delta _j}dtf\left( t\right) \cos k\left( t-t_{j-1}\right) -\frac
1{\sin k\left| \Delta _{j+1}\right| }\int_{\Delta _j+1}dtf\left( t\right)
\cos k\left( t-t_j\right) \right] \right|  
$$
we expand $F(t_{j-1})=\int_{t_{j-1}}^{t_j}dt\,f(t)\cos k\left(
t-t_{j-1}\right) $ and $G(t_{j+1})=\int_{t_j}^{t_{j+1}}dt\,f(t)\cos k\left(
t-t_j\right) $ around $t_j.$ This yields with $\eta _j\in \Delta _j$ and $%
\eta _{j+1}\in \Delta _{j+1}$ 
$$
\leq k\left| z\right| X\left| \sum_{j=1}^nf\left( t_j\right) \left[ \frac{%
\left| \Delta _j\right| }{\sin k\left| \Delta _j\right| }-\frac{\left|
\Delta _{j+1}\right| }{\sin k\left| \Delta _{j+1}\right| }\right] \right| + 
$$
$$
+k\left| z\right| X\sum_{j=1}^n\left[ \frac{\left( t_{j-1}-t_j\right) ^2}{%
2\sin k\left| \Delta _j\right| }\left( -f^{\prime }\left( \eta _j\right)
-k^2\int_{\eta _j}^{t_j}dtf\left( t\right) \cos k\left( t-\eta _j\right)
\right) \right] - 
$$
$$
-k\left| z\right| X\sum_{j=1}^n\left[ -\frac{\left( t_{j+1}-t_j\right) ^2}{%
2\sin k\left| \Delta _{j+1}\right| }\left( f^{\prime }\left( \eta
_{j+1}\right) \cos k\left( \eta _{j+1}-t_j\right) -kf\left( \eta
_{j+1}\right) \sin k\left( \eta _{j+1}-t_j\right) \right) \right]  
$$
Since $\stackunder{0\leq x\leq \frac \pi 2}{\sup }\left( \frac x{\sin
x}\right) ^{\prime }=1$ then the first term above is bounded by\vspace{-3mm}%
$$
2k\left| \Delta \right| \left| z\right| X\left\| f\right\|  
$$
For the second term we obtain the bound%
$$
\left| z\right| X\frac{\left| \Delta \right| }4\sup _\Delta \left| f^{\prime
}\right| +\frac{k^2\left| \Delta \right| ^2}4\sup _\Delta \left| f\right|
+\frac \pi 4\left| \Delta \right| \sup _\Delta \left| f^{\prime }\right| +%
\frac{\pi k}4\left| \Delta \right| \sup _\Delta \left| f\right| \leq  
$$
$$
\leq \frac{\left| z\right| X\left| \Delta \right| \pi }4\left( \left\|
f\right\| \left( 2+k^2\left| \Delta \right| +k\right) \right)  
$$
Putting all of this together we finally arrive at 
$$
\Big|T\Big(I_h\tprod\limits_{j=1}^n\delta \left( B(t_j)-x_j\right) \Big)%
\left( zf\right) \Big|\leq \left( {\textstyle \prod\limits_{j=1}^{n+1}\sqrt{%
\frac 1{4\left| \Delta _j\right| }}}\right) \exp \left( L\left| z\right|
X\left\| f\right\| +\left( \frac \pi 2\left| \Delta \right| +\frac 12\right)
\left| z\right| ^2\left\| f\right\| ^2\right)  
$$
where $L=\pi +\frac 34\pi k\left| \Delta \right| +2k\left| \Delta \right|
+\frac \pi 4\left| \Delta \right| \left( 2+k^2\left| \Delta \right| \right) $
is a constant.

Hence we have the following estimate 
\begin{equation}
\label{Main}\Big|T\Big(I_h\tprod\limits_{j=1}^n\delta \left(
B(t_j)-x_j\right) \Big)\left( zf\right) \Big|\leq \left( {\textstyle %
\prod\limits_{j=1}^{n+1}\sqrt{\frac 1{4\left| \Delta _j\right| }}}\right)
\exp \left( X^2\gamma \right) \exp \left[ \left| z\right| ^2\left\|
f\right\| ^2\left( \frac 12+\frac \pi 2\left| \Delta \right| +\frac{L^2}{%
2\gamma }\right) \right] 
\end{equation}
where $\gamma >0.$

\noindent {\bf 2.\ part:} In this final step we use the method developed in 
\cite{LLSW} to control the convergence of (\ref{result}). Although the
slight modification to our case is easy we give the basic steps for the
convenience of the reader.

In order to apply theorem 2.3 to perform the integration we need to show that%
$$
\left( {\textstyle \prod\limits_{j=1}^{n+1}\sqrt{\frac 1{4\left| \Delta
_j\right| }}}\right) \exp \left( X^2\gamma \right) 
$$
is integrable with respect to $\nu .$ To this end we choose $q>2$ and $%
0<\gamma <\frac \beta q.$ With this choice of $\gamma $ the property i) of $%
\nu $ yields that $\exp \left( \gamma X^2\right) \in L^q\left( {\bf R}%
^n\times \Lambda _n,\left| \nu \right| \right) $ and with%
$$
Q\equiv \left( \dint_{{\bf R}}\dint_\Delta \left| \nu \right| \left(
dx,dt\right) \exp \left( \gamma qx^2\right) \right) ^{\frac 1q} 
$$
we have%
$$
\left( \dint_{{\bf R^{\mit n}}}\dint_{\Lambda _n}\tprod\limits_{j=1}^n\left|
\nu \right| \left( dx_j,dt_j\right) \exp \left( \gamma qX^2\right) \right)
^{\frac 1q}\leq \exp \left( \gamma \left( x_o^2+x^2\right) \right)
Q^n<\infty . 
$$

Now we choose $p$ such that $\frac 1p+\frac 1q=1.$ Using the property ii) of 
$\nu $ and the formula%
$$
\int_{\Lambda _n}d^nt\tprod\limits_{j=1}^{n+1}\left( \frac 1{4\left|
t_j-t_{j-1}\right| }\right) ^\alpha =\left( \frac{\Gamma \left( 1-\alpha
\right) }{4^\alpha }\right) ^{n+1}\frac{\left| \Delta \right| ^{n\left(
1-\alpha \right) -\alpha }}{\Gamma \left( \left( n+1\right) \left( 1-\alpha
\right) \right) },\text{ }\alpha <1 
$$
we obtain the following bound%
$$
\left[ \int_{{\bf R^{\mit n}}}\int_{\Lambda _n}\tprod\limits_{j=1}^n\left|
\nu \right| \left( dx_j,dt_j\right) \tprod\limits_{j=1}^{n+1}\left( \frac
1{4\left| t_j-t_{j-1}\right| }\right) ^{\frac p2}\right] ^{\frac 1p}\leq  
$$
$$
\leq \left| \nu \right| _{t\infty }^{n/p}\frac{\Gamma \left( {\textstyle%
\frac{2-p}2}\right) ^{\frac{n+1}p}\left| \Delta \right| ^{\frac np-\frac
12\left( n+1\right) }}{4^{\frac{n+1}2}\Gamma \left( \left( n+1\right) {%
\textstyle\frac{2-p}2}\right) ^{\frac 1p}}<\infty  
$$
$\left| \nu \right| _{t\infty }$ is shorthand notation for the essential
supremum of the $L^\infty $-density of $|\nu |_t$ which exists due to
condition {\it ii}).

Finally an application of H\"older's inequality gives%
$$
\left| \left( \tprod\limits_{j=1}^{n+1}\sqrt{\textstyle \frac 1{4\left|
t_j-t_{j-1}\right| }}\right) \exp \left( \gamma X^2\right) \right| _1\leq 
$$
$$
\leq \exp \left( \gamma x_o^2+\gamma x^2\right) Q^n\left| \nu \right|
_{t\infty }^{n/p}\frac{\Gamma \left( {\textstyle \frac{2-p}2}\right) ^{\frac{%
n+1}p}\left| \Delta \right| ^{\frac np-\frac 12\left( n+1\right) }}{%
2^{n+1}\Gamma \left( \left( n+1\right){\textstyle \frac{2-p}2} \right)
^{\frac 1p}}\equiv C_n<\infty 
$$
Hence theorem 2.3 yields%
$$
I_n\equiv \int_{{\bf R^{\mit n}}}\int_{\Lambda _n}\tprod\limits_{j=1}^n\nu
\left( dx_j,dt_j\right) \left( I_h\prod_{j=1}^n\delta \left( B\left(
t_j\right) -x_j\right) \right) \in \left( {\cal S}\right) ^{*}. 
$$

As the $C_n$ are rapidly decreasing in $n$ the hypotheses of theorem 2.2 are
fulfilled and hence

$$
I_V=\dsum\limits_{n=0}^\infty I_n\in \left( {\cal S}\right) ^{*}\text{ .} 
$$
\TeXButton{End Proof}{\endproof}

{\bf Acknowledgements}

It is a pleasure to thank Professor L. Streit for initiating this
collaboration and his constant encouragement and advice. We thank Professors
M. de Faria, D. C. Khandekar and Yu.\ G. Kondratiev for helpful discussions.
We are grateful to STRIDE\footnote{%
Program of Junta Nacional de Investigacao Cientifica e Technologica
(Portugal)}, whose generous support made this collaboration possible.

\TeXButton{TeX field}{\small}


\begin{thebibliography}{99}
\bibitem{AHK}  Albeverio, S.A., H\NEG oegh-Krohn, R. (1976), {\it %
Mathematical Theory of Feynman Integrals}. LNM {\bf 523}, Springer Verlag.

\bibitem{Ex}  Exner, P. (1985), {\it Open Quantum Systems and Feynman
Integrals.} Reidel, Dordrecht.

\bibitem{FPS}  Faria, M. de, Potthoff, J. and Streit, L. (1991), {\it The
Feynman Integrand as a Hida Distribution}. J. Math. Phys. {\bf 32}, 2123.

\bibitem{H}  Hida, T. (1980), {\it Brownian Motion,} Applications of
Mathematics {\bf 11}, Springer Verlag, Berlin.

\bibitem{HKPS}  Hida, T., Kuo, H.H., Potthoff, J. and Streit, L. (1993), 
{\it White Noise: An Infinite Dimensional Calculus}. Kluwer Academic
Publishers, Dordrecht.

\bibitem{HS}  Hida, T. and Streit, L. (1983), {\it Generalized Brownian
Functionals and the Feynman Integral. }Stoch. Proc. Appl. {\bf 16}, 55.

\bibitem{Ho}  Holstein, B.R. (1992), {\it Topics in advanced Quantum
mechanics.} Redwood City, Ca., Addison-Wesley.

\bibitem{KS}  Khandekar, D.C. and Streit, L.(1992), {\it Constructing the
Feynman Integrand.} Ann. Physik {\bf 1}, 49-55.

\bibitem{KoS}  Kondratiev, Yu.G. and Streit, L. (1992), {\it Spaces of White
Noise Distributions}. {\it Constructions, Descriptions, Applications I}. To
appear in Rep. Math. Phys. {\bf 33}.

\bibitem{KLPSW}  Kondratiev, Yu.G., Leukert, P., Potthoff, J., Streit, L.,
Westerkamp, W. (1994), {\it Generalized Functionals in Gaussian Spaces - The
Characterization Theorem Revisited -} Preprint Univ. Mannheim Nr. 175/94.

\bibitem{KT}  Kubo, I. and Takenaka, S. (1980), {\it Calculus on Gaussian
White Noise I+II}. Proc. Japan Acad. 56A, 376-380 and 411-416.

\bibitem{Kuo}  Kuo, H.H. (1991), {\it Lectures on White Noise Analysis}.
Soochow J. Math. {\bf 18} 229-300.

\bibitem{LLSW}  Lascheck, A., Leukert, P., Streit, L. and Westerkamp, W.
(1993), {\it Quantum Mechanical Propagators in Terms of Hida Distributions}.
Rep. Math. Phys. {\bf 33}, 221-232.

\bibitem{MY}  Meyer, P.A. and Yan, J.--A. (1990), {\it Les ''fontions
caract\'eristiques'' des distribuitions sur l'\'espace de Wiener.} Seminaire
de Probabilites XXV, ed.: P. A. Meyer, M. Yor, Springer, p. 61-78.

\bibitem{Ob}  Obata, N. (1994), {\it White Noise Calculus and Fock Space.}
LNM {\bf 1577}. Springer, Berlin.

\bibitem{PS}  Potthoff, J. and Streit, L. (1991), {\it A Characterization of
Hida Distributions}. J. Funct. Anal. {\bf 101}, 212-229.

\bibitem{P}  Potthoff, J. (1991), {\it Introduction to White Noise Analysis}%
. Baton Rouge Preprint.

\bibitem{S1}  Streit, L. (1993), {\it The Feynman Integral - Recent Results. 
}In: Dynamics of complex and Irregular Systems. Ph. Blanchard et al., eds.
World Scientific.

\bibitem{S2}  Streit, L. (1994), {\it White Noise Analysis and Applications
in Quantum Physics}. In: Stochastic Analysis and Applications in Physics.
Ed.: A.I. Cardoso et al.; Kluwer Dordrecht, in print.

\bibitem{SW}  Streit, L. and Westerkamp, W. (1993), {\it A generalization of
the characterization theorem for generalized functionals of White Noise.}
In: Dynamics of complex and Irregular Systems. Ph. Blanchard et al., eds.
World Scientific.

\bibitem{W}  Westerkamp, W. (1993), {\it A Primer in White Noise Analysis}.
In: Dynamics of complex and Irregular Systems. Ph. Blanchard et al., eds.
World Scientific.

%
%
%
%
\end{thebibliography}
\end{document}